\documentstyle[preprint,aps]{revtex}

\begin{document}
\draft

\title{A Toy Model for Blandford-Znajek Mechanism}

\author{Li-Xin Li}
\address{Department of Astrophysical Sciences, Princeton University, Princeton, 
NJ 08544}
\date{February 24, 1999; Revised December 8, 1999}
\maketitle

\begin{abstract}

A toy model for the Blandford-Znajek mechanism is investigated: 
a Kerr black hole with a toroidal electric current residing in a
thin disk around the black hole. The toroidal electric
current generates a poloidal magnetic field threading the black hole and
disk. Due to the interaction of the magnetic field with remote charged particles,
the rotation of the black hole and disk 
induces an electromotive force, which can power an
astrophysical load at remote distance. The power
of the black hole and disk is calculated. It is found that, for a wide range of 
parameters specifying the rotation of the black hole and
the distribution of the electric current in the disk,
the power of the disk exceeds  the power of the black hole.
The torque provided by the black hole and disk is also calculated. 
The torque of the disk is comparable to the torque of the black hole.
As the disk loses its angular momentum, the mass of the disk gradually drifts
towards the black hole and gets accreted. 
Ultimately the power comes from the gravitational binding energy between
the disk and the black hole, as in the standard theory of accretion disk, 
instead of the rotational energy of the black hole. 
This suggests that the Blandford-Znajek mechanism may be less efficient in 
extracting energy from a rotating black hole with a thin disk.
The limitations of our simple model and possible improvements deserved
for future work are also discussed.

\end{abstract}


\section{Introduction}

It's well believed that black holes exist in many astrophysical systems, 
such as in active galactic nuclei (AGNs), centers of galaxies,
and some stellar binary systems. For 
a rotating black hole with an accretion disk, magnetic field threading the black 
hole and the disk could exist (\cite{tho86} and references
therein). Due to the rotation of the black hole relative to the magnetic
field, the black hole's rotational energy can be 
extracted through the Blandford-Znajek mechanism \cite{bla77}. 
Currents in the disk are needed
to confine the magnetic field, but without using any mass accretion 
we can tap the huge rotational energy of the black hole. The Blandford-Znajek
process provides a mechanism for extraction of energy from a rotating black
hole which is more promising than the Penrose process in practice 
\cite{pen69,car79}. For a long time the Blandford-Znajek mechanism
has been considered as a reasonable process powering the radio jets 
in AGNs \cite{ree82,beg84}. Recently, this mechanism has 
been invoked in models for gamma-ray bursts (GRBs) where a rotating  black 
hole with an accretion disk (or torus) forms through the collapse of a 
rotating massive star or the merger of a black hole with a neutron star
\cite{pac93,woo93,pac98,lee99}. The Blandford-Znajek mechanism
is favorable for these phenomena since ultra-relativistic Lorentz factors
are required for jets in AGNs and GRBs and it's believed that very clean
energy can be created through the Blandford-Znajek mechanism \cite{bla99}.

However, though some numerical calculations have been taken (\cite{mac84},
\cite{gho97} and references therein), by now there are no clear
answers to such questions as how the magnetic field is generated,
how it is distributed, how much energy can be extracted from the 
black hole and disk, what portion of the extracted energy is attributable
to the black hole, and if the energy extracted from the black hole is
clean enough. Realistic cases are too complicated, we
have very little observational clues. 

In this paper, instead of searching  for complicated numerical solutions, we
consider a semi-analytical toy model to probe the Blandford-Znajek mechanism. The
toy model is a Kerr black hole with a toroidal electric current residing in a
geometrically thin disk around the Kerr black hole. The toroidal current 
generates a poloidal
magnetic field threading the black hole and disk. Due to the interaction
between the magnetic field and remote charged particles, the rotation of the
black hole and disk induces an electromotive force (EMF) on the
black hole's horizon and on the disk. This EMF could be the energy source for
remote astrophysical loads (such as the jets in AGNs and GRBs). The 
power and torque provided by the black hole and disk are calculated. 
It is found that for a wide range of parameters specifying the rotation
of the black hole and the distribution of the
current density in the disk, the power provided by the disk exceeds 
the power provided by the black hole. The torque provided by the disk is
comparable to the torque provided by the black hole. 
This agrees with the results in \cite{bla77,liv98} and suggests that the 
Blandford-Znajek mechanism may be less efficient in extracting energy from a 
rotating black hole for the thin disk case.

Though the case of thin disks is simple for calculation, in real astronomy
accretion disks may be geometrically thick and the case of thick disks may
be quite different form that of thin disks\cite{ree82,bla99,arm99}. 
However the simple model presented
in the paper gives a complete semi-analytical example: from the generation
of magnetic fields to the extraction of energy from the black hole and
the disk. The extension to the case of thick disks is extremely interesting
and challenging.

\section{Description of the model}
Realistic cases of a rotating black hole with a disk and magnetic field are
likely to be extremely complicated. However, 
if the the magnetic field associated with the black hole has a somewhat
poloidal structure, it can be modeled as being generated by some toroidal 
electric current outside the black hole's horizon. This toroidal electric 
current most likely resides in the disk around the black hole. 
Thus here we consider a model of a Kerr black hole with a thin disk in the 
equatorial plane and there is a distribution of electric current in
the disk. The Kerr black hole has mass $M$, angular momentum $Ma$
[throughout the paper we use the geometric units with $G = c =1$ and the 
Boyer-Lindquist coordinates $(t,r,\theta,\phi)$ for Kerr black hole]. Then the 
angular velocity of the black hole's horizon is
\begin{eqnarray}
    \Omega_H={a\over 2Mr_H},
\end{eqnarray}
where $r_H=M+\sqrt{M^2-a^2}$ is the radius of the Kerr black hole's
outer horizon ($a^2\leq M^2$). As usual the disk's angular velocity is taken to be 
the relativistic Keplerian angular velocity \cite{nov73}
\begin{eqnarray}
    \Omega_D(r)=\left({M\over r^3}\right)^{1/2}{1\over 1+a\left({M\over r^3}
             \right)^{1/2}}.
\end{eqnarray}
The outer edge of the disk is at $r=r_b$, the inner edge of the disk is taken
to be at the innermost stable circular orbit (the ``marginally stable" orbit) in
the equatorial plane \cite{nov73}:
\begin{eqnarray}
    r_{\rm ms}=M\left\{3+z_2-\left[(3-z_1)(3+z_1+2z_2)\right]^{1/2}\right\},
\end{eqnarray}
where
$z_1=1+(1-a^2/M^2)^{1/3}\left[(1+a/M)^{1/3}+(1-a/M)^{1/3}\right]$ and
$z_2=\left(3a^2/M^2+z_1^2\right)^{1/2}$. (The assumption of a Keplerian
disk is valid only if the magnetic field is weak enough \cite{kro99,mei99}.)
The toroidal electric current in the disk has a surface density $J=J(r)$
which is distributed between $r_{ms}$ and $r_b>r_{ms}$.

The magnetic field generated by a single toroidal electric
current at a fixed radius [i.e. the current density $J(r)$ is a delta function] has
been well investigated by many authors \cite{zna78,pet75,lin79}. By linear superposition,
the magnetic flux through a surface bounded by a circle with $r={\rm const}$
and $\theta={\rm const}$ is
\begin{eqnarray}
    \Psi(r,\theta)=2\pi A_\phi(r,\theta)=2\pi\int_{r_{\rm ms}}^{r_b}
    J(r^\prime){dA_\phi\over dr^\prime} dr^\prime,
\end{eqnarray}
where $A_\phi$ is the toroidal component of the electric vector potential, 
and $dA_\phi/dr^\prime$ is 
\begin{eqnarray}
    {dA_\phi\over dr^\prime} &=& 2\sum_{l=1}^\infty
    \left\{\alpha_l^r\left[ra\sin^2\theta{\Delta\over
    \Sigma}{1\over\sqrt{M^2-a^2}}P_l^\prime(u)P_l(\cos\theta)-a\sin^2\theta\cos\theta
    {r^2+a^2\over\Sigma}P_l(u)P_l^\prime(\cos\theta)\right]\right.\nonumber\\
     &&+\alpha_l^i\left[-a^2\sin^2\theta\cos\theta{\Delta\over\Sigma}{1\over
     \sqrt{M^2-a^2}}P_l^\prime(u)P_l(\cos\theta)-r\sin^2\theta{r^2+a^2\over\Sigma}
     P_l(u)P_l^\prime(\cos\theta)\right.\nonumber\\
     &&\left.\left.+{\Delta\sin^2\theta\over l(l+1)}{1\over\sqrt{
     M^2-a^2}}P_l^\prime(u)P_l^\prime(\cos\theta)\right]\right\}\nonumber\\
     &&+2\sum_{l=1}^\infty\left\{\beta_l^r\left[ra\sin^2\theta{\Delta\over
    \Sigma}{1\over\sqrt{M^2-a^2}}Q_l^\prime(u)P_l(\cos\theta)-a\sin^2\theta\cos\theta
    {r^2+a^2\over\Sigma}Q_l(u)P_l^\prime(\cos\theta)\right]\right.\nonumber\\
     &&+\beta_l^i\left[-a^2\sin^2\theta\cos\theta{\Delta\over\Sigma}{1\over
     \sqrt{M^2-a^2}}Q_l^\prime(u)P_l(\cos\theta)-r\sin^2\theta{r^2+a^2\over\Sigma}
     Q_l(u)P_l^\prime(\cos\theta)\right.\nonumber\\
     &&\left.\left.+{\Delta\sin^2\theta\over l(l+1)}{1\over\sqrt{
     M^2-a^2}}Q_l^\prime(u)P_l^\prime(\cos\theta)\right]\right\},
\end{eqnarray}
where $\Delta = r^2-2Mr+a^2$, $\Sigma=r^2+a^2\cos^2\theta$,
$A=(r^2+a^2)^2-\Delta a^2 \sin^2\theta$, $u = (r-M)/\sqrt{M^2-a^2}$, $P_l(z)$
and $Q_l(z)$ are Legendre functions, and $P_l^\prime(z) = dP_l(z)/dz$,
$Q_l^\prime(z) = dQ_l(z)/dz$, and
the coefficients $\alpha_l^r$, $\alpha_l^i$, $\beta_l^r$, 
and $\beta_l^i$ are respectively \\
(1) for $r<r^\prime$, $\beta_l^r=\beta_l^i=0$ for all $l$; but
\begin{eqnarray}
    \alpha_l^r={(2l+1)\pi\over l(l+1)(M^2-a^2)}\left(\Sigma^\prime\over A^\prime\right)^{1/2}
     \Delta^\prime a P_l(0) Q_l^\prime(u^\prime),
\end{eqnarray}
\begin{eqnarray}
    \alpha_l^i&=&{(2l+1)\pi\over l(l+1)\sqrt{M^2-a^2}}\left(\Sigma^\prime\over A^\prime\right)^{1/2}
    \left[-({r^\prime}^2+a^2)P_l^\prime(0)Q_l(u^\prime)\right.\nonumber\\
    &&\left.+{\Sigma^\prime\Delta^\prime\over
    r^\prime l(l+1)}{1\over\sqrt{M^2-a^2}}P_l^\prime(0)Q_l^\prime(u^\prime)\right];
\end{eqnarray}
(2) for $r>r^\prime$, $\alpha_l^r=\alpha_l^i=0$ for all $l$, but
\begin{eqnarray}
    \beta_l^r={(2l+1)\pi\over l(l+1)(M^2-a^2)}\left(\Sigma^\prime\over A^\prime\right)^{1/2}
     \Delta^\prime a P_l(0) P_l^\prime(u^\prime),
\end{eqnarray}
\begin{eqnarray}
    \beta_l^i&=&{(2l+1)\pi\over l(l+1)\sqrt{M^2-a^2}}\left(\Sigma^\prime\over 
    A^\prime\right)^{1/2}
    \left[-({r^\prime}^2+a^2)P_l^\prime(0)P_l(u^\prime)\right.\nonumber\\
    &&\left.+{\Sigma^\prime\Delta^\prime\over
    r^\prime l(l+1)}{1\over\sqrt{M^2-a^2}}P_l^\prime(0)P_l^\prime(u^\prime)\right];
\end{eqnarray}
where $\Delta^\prime=\Delta(r=r^\prime)$,
$\Sigma^\prime=\Sigma(r=r^\prime,\theta=\pi/2)={r^\prime}^2$, and 
$A^\prime=A(r=r^\prime,\theta=\pi/2)$.
The normal component of the magnetic field in the disk is
\begin{eqnarray}
    B_D = {1\over 2\pi} \left({\Delta\over A}\right)_{\theta={\pi\over2}}^{1/2}
        {d\Psi(r,\pi/2)\over dr}.
\end{eqnarray}
The poloidal magnetic field on the horizon (which is perpendicular to the horizon) is
\begin{eqnarray}
    B_H = {1\over 2\pi\left(r_H^2+a^2\right)\sin\theta}{d\Psi(r_H,\theta)
        \over d\theta}.
\end{eqnarray}

Due to the interaction of the magnetic field with remote charged particles, 
the rotation of the black hole and disk induces EMFs on the horizon and in the disk \cite{mac82}. The total EMF on the black hole is 
\begin{eqnarray}
    {\cal E}_H={1\over 2\pi}\Omega_H\Psi(r_H),
\end{eqnarray}
where $\Psi(r_H) = \Psi(r_H, \pi/2)$ is the magnetic flux through the northern
hemi-sphere of black hole's horizon. The total EMF in the disk is
\begin{eqnarray}
    {\cal E}_D={1\over 2\pi}\int_{r_{ms}}^{r_c}\Omega_D d\Psi(r)
    ={1\over 2\pi}\left[\Omega_D(r_c)
    \Psi\left(r_c\right)-\Omega_D(r_{\rm ms})
    \Psi\left(r_{\rm ms}\right)-\int_{r_{\rm ms}}^{r_c}
    \Psi{d\Omega_D\over dr}dr\right],
\end{eqnarray}
where $\Psi(r) = \Psi(r,\pi/2)$, $r_c < r_b$ is the radius within 
which the disk's energy is available. (As in the case of the Sun, the open magnetic
field lines may emerge from only a small fraction of the disk surface 
\cite{bla82}. So we should choose $r_c\ll r_b$ in practice.)
These EMFs could be the energy source powering a remote astrophysical load. 
The black hole, the disk, and the remote astrophysical load form an electric 
circuit in series. (Alternative type of circuits are possible, see the discussions
in Sec.~\ref{sec4}.) The circuit has two batteries --- one is the Kerr
black hole with EMF ${\cal E}_H$ and internal resistance $Z_H$ (which is
of several hundred ohms), 
the other is the disk with EMF ${\cal E}_D$ and  negligible
internal resistance (i.e. the disk's resistance is supposed to
be $\ll Z_H$) --- and a resistor which is the astrophysical load
with resistance $Z_A\equiv\alpha Z_H$. Suppose ${\cal E}_D$ and
${\cal E}_H$ generate a (single) poloidal electric current $I=\left(
{\cal E}_H+{\cal E}_D\right)/\left(Z_H+Z_A\right)$, which flows along
the circuit from the disk to the black hole, from the black hole up to the
remote astrophysical load along the symmetry axis, from the remote astrophysical 
load down to the disk at the circle with radius $r_c$ ($r_{ms}<r_c < r_b$) 
in the disk. The power provided to the remote astrophysical load, 
$P=I^2 Z_A$, is sensitive to the ratio $\alpha = Z_A/Z_H$. $P$ reaches its 
maximum at $\alpha =1$. $\alpha = 1$ is called the impedance matching
condition \cite{mac82}.
Define $\xi={\cal E}_D/{\cal E}_H$, the ratio of the power of the
disk $P_D\equiv-dE_D/dt=I{\cal E}_D$ to the effective power of the black hole 
$P_H\equiv-dE_H/dt=I{\cal E}_H-I^2 Z_H$
(where $E_D$ is the energy of the disk, $E_H$ is the energy of the black hole; 
$P_D +P_H = P$) is
\begin{eqnarray}
    {P_D\over P_H}= {(1+\alpha)\xi\over \alpha-\xi}.
    \label{pdh}
\end{eqnarray}
If $\xi\ge\alpha$, the power of the black hole is negative or
zero (while the power of the disk is always positive).
If $\xi_{\rm cr}<\xi<\alpha$, where $\xi_{\rm cr} \equiv \alpha/(2+\alpha)$, 
the power of the black hole is positive but less
than the power of the disk, i.e. $0<P_H<P_D$.
If $\xi\le\xi_{\rm cr}$, the power of the black hole is bigger
than the power of the disk. Thus, the parameter $\xi$
gives a sensible measure of the relative importance of the disk's power
and the black hole's power. If $\xi>\xi_{\rm cr}$, the disk's power 
dominates. From the definition of $\xi_{\rm cr}$, $\xi_{\rm cr}<1$ always. 
(For the case of $P=P_m$ i.e. $\alpha=1$, we have $\xi_{\rm cr}=1/3$.) 
Therefore, if $\xi>1$, we can conclude that the disk's power always 
dominates the black hole's power.

The torque produced by the black hole is
\begin{eqnarray}
    T_H\equiv -{dL_H\over dt}={I\over 2\pi}\Psi\left(r_H\right),
\end{eqnarray}
where $L_H=Ma$ is the angular momentum of the black hole.
The torque produced by the disk is
\begin{eqnarray}
    T_D\equiv -{dL_D\over dt}={I\over 2\pi}\left[\Psi\left(r_c\right)-
    \Psi\left(r_{\rm ms}\right)\right],
\end{eqnarray}
where $L_D$ is the angular momentum of the disk,
$\Psi(r_c)-\Psi(r_{\rm ms})$ is the magnetic flux through the 
disk inside the circle with $r=r_c$. The ratio of $T_D$ to $T_H$ is
\begin{eqnarray}
    \eta\equiv{T_D\over T_H}={\Psi(r_c)-\Psi(r_{\rm ms})\over \Psi(r_H)}.
\end{eqnarray}

\section{Solutions of the model}
If we do not care about the absolute values of the power and torque of 
the black hole and disk, the problem is ``self-similar'' in the sense 
that the mass of the black hole and the absolute magnitude of the magnetic field 
are not important for us. Then, we are left with three adjustable
parameters: (1) $a/M$, where $M$ is the mass of the black hole and
$Ma$ is the angular momentum of the black hole; (2) the shape of the 
surface current function $J=J(r)$; and (3) $r_c/M$, where $r_c$
is the radius within which the disk's energy is available. 
We take $J=J_0 (r/r_{ms})^{-n}$,
where $n$ is a positive dimensionless number, $J_0$ is the magnitude of $J$
at the inner edge of the disk. The radius $r_c$ is usually thought to be 
several times larger than $r_H$ \cite{liv98} but much smaller than $r_b$ 
\cite{bla82}. Thus we take $r_c =r_{ms}+\kappa r_H$, 
where $\kappa$ is a positive constant which has magnitude $\sim 1$.
Then the three adjustable dimensionless parameters are $a/M$, $n$, and $\kappa$
(the value of $J_0$ is unimportant for us). Clearly, if the magnetic field
in the disk, $B_D$, keeps the same sign everywhere (no reversal of magnetic 
field), the larger the radius $r_c$ is, larger the power of the disk is, and
thus less efficient the Blandford-Znajek mechanism is. To avoid overestimating
the power of the disk by choosing a large $r_c$, in our numerical solutions
we take $\kappa =0.2$. Even for such a small value of $\kappa$, we will see that
the disk's power still dominates the black hole's power. Thus,
our numerical calculation is
taken for $\kappa=0.2$ and a series of values of $a/M$ from $0.1$ to $0.99$,
$n$ from $1$ to $3$ (For a stationary Newtonian disk we have $n=1$ \cite{bla76}). 
(For $n\ge 1$, the results are insensitive to the value of 
$r_b/M$ for sufficiently large $r_b$.)

During the numerical calculation, the summation $\sum_l$ over Legendre functions 
\cite{zna78} is truncated at $l_{\rm max} = 10$. [For a realistic disk
with thickness $\Delta\theta$, the summation over $l$ should be
truncated at $l_{\rm max}\approx (\Delta\theta)^{-1}$.] 
The results are shown in Fig.~\ref{figure1} -- Fig.~\ref{figure4}. The results 
are insensitive to $n$. 
Fig.~\ref{figure1} shows the dimensionless parameter $\xi={\cal E}_D/{\cal E}_H$, 
which measures the relative importance of the power of the disk to the 
power of the black hole. For the impedance matching case with $Z_A=Z_H$,  
the power of the disk always dominates the power of the black hole. For example, for
$a/M=0.9$ and $n=2$, we have $\xi={\cal E}_D/{\cal E}_H =0.68$, $P_D/P_H=4.3$
[Eq.~(\ref{pdh}) with $\alpha=1$]. Fig.~\ref{figure2} shows the ratio $P_D/P_H$ 
for the case with $n=2$ and $Z_A =Z_H$. For $a/M<0.54$, the black hole's effective 
power $P_H$ is negative
(i.e. $I^2 Z_H> {\cal E}_H$; $P_D$ is always positive since we have assumed 
$Z_D=0$); for $a/M>0.54$, both $P_H$ and $P_D$ are positive but $P_D>P_H$. Thus
the disk's power dominates the black hole's power.
Fig.~\ref{figure3} shows the dimensionless parameter $\eta=T_D/T_H$, 
which measures the relative importance of the torque of the disk to the torque 
of the black hole. The torque of the disk is comparable
to the torque of the black hole. Fig.~\ref{figure4} shows the normal components
of the poloidal magnetic field on the horizon and disk, which use 
$B_0=J_0/c$ as unit. The magnetic field is most strong at the inner edge of the 
disk. 

\section{Discussions and conclusions}
\label{sec4}
From the simple model presented in the paper we have got some insight into
the Blandford-Znajek mechanism. Since the electric
current generating the magnetic field must reside outside the black hole and
most likely resides in a disk around the black hole, as described in our thin 
disk model, the strength of the magnetic field at the inner edge of the disk 
should be bigger than that on the horizon \cite{gho97,liv98}. The
solutions of our model show that this is true. Though the average strength
of the magnetic field in the whole disk could be smaller than that on the horizon, the
local strength of the magnetic field at the inner edge of the disk could be
much stronger. In fact, in our solutions $B_D(r=r_{ms})/B_H(\theta=0)$ has never
got smaller than $1$. Our solutions show that, for a wide range of parameters
specifying the model considered in the paper, the
power of the disk is stronger than the power of the black hole.  
The torque provided by the disk is comparable to the torque provided by the 
black hole. These results are physically plausible since (1) the electric 
current generating the magnetic field resides in the disk and thus
the magnetic field at the inner edge of the disk (where the current density
is most strong) should be stronger than that on the black hole's horizon; 
(2) rapidly rotating black holes exclude stationary and axisymmetric magnetic fields
\cite{bic85}, as seen from Fig.~\ref{figure4} (left diagram);
(3) the disk has a surface area larger than that of the horizon and thus
the magnetic flux through the disk could be larger than that through
the black hole; (4) the disk is a perfect conductor with a negligible resistance 
while the black hole's horizon has a significant internal resistance of several hundred 
ohms which consumes the power of the black hole and disk. 

For the model considered in the paper the angular momentum and energy lost from
the disk are important. As the disk loses its angular momentum, the mass of the disk 
gradually drifts towards the black hole and gets accreted. Ultimately the power at the
remote load comes from the gravitational binding energy between
the disk and the black hole, as in the standard theory of accretion disk, 
instead of the rotational energy of the black hole. 
This implies that the Blandford-Znajek mechanism is not likely to be efficient in
extracting energy from a rotating black hole with a geometrically thin disk 
and magnetic field.

Though the simple model is simple and convenient for calculations and we believe 
that some essential features for the Blandford-Znajek mechanism have been included, 
the limitations must also be emphasized: (1) We have assumed that the disk is 
geometrically thin. But in real cases this may not be true, especially for subcritical 
accretion. There could be an alternative type of models where, in the limiting case, 
poloidal flux is confined by and excluded from a funnel formed by an ion torus with
ADAF or ADIOS solution where the disk's power is manifestly zero 
\cite{ree82,bla99,bla99b}.
(2) We have neglected the magnetic coupling between the black hole and the disk. In
fact there could be strong coupling between the disk and the hole with the former
acquiring energy from the work done by the later \cite{tho86,bla99b,gam99}, which
might lead that some of the power of the disk effectively comes from the
rotation of the black hole \cite{mei99}. In particular, loops of closed lines 
connecting the disk with the hole could exist and play
important roles in transportation of angular momentum and energy.
For a fast rotating black hole ($a/M>0.36$) with a relativistic Keplerian
disk, the rotation of the disk is slower than the rotation of the black hole.
So the closed magnetic field lines will transfer
angular momentum from the hole to the disk and reverse the accretion flow
\cite{bla99b}. Accompanying the angular momentum, energy will also be transfered
from the black hole to the disk via Poynting flux. Then for the remote load
the energy directly extracted from the black hole is decreased, but it may be
compensated by the energy extracted from the disk. We can imagine that with
suitable conditions a steady state could exist when the power from the disk
balances the energy flux from the black hole to the disk and then all the power
of the disk effectively comes from the spin of the black hole. (3) We
have supposed that the poloidal magnetic field is generated purely by the toroidal
currents in the disk. By doing so we have neglected the effect of the currents
induced by the charges in the magnetosphere. This allows to calculate
the poloidal magnetic field from the potential instead of solving the 
complicated Grad-Shafronov equation. With this approach the back-reaction
of the induced charges and currents cannot be taken into account, but it is particularly
useful for semi-analytical investigations. (4) We have modeled the toroidal
magnetic field with a global poloidal current loop flowing through the black hole,
the accretion disk, and the remote load. This allows us to conveniently compare 
the power of the black hole and the power of the disk since they have the
same unique current. The realistic case will be more complicated since poloidal
currents flow into and out of the black hole and the disk diversely at all radius.
An alternative type of current loops is that the poloidal current associated
with black hole and the poloidal current associated with the disk flow
separately, i.e. we have two distinct poloidal current loops. In this case
the black hole and the disk have different loads and it's hard to compare
their powers. However if the disk and the black hole have comparable EMFs
and the impedance matching conditions are satisfied for both the hole's circuit
and the disk's circuit, the power of the disk will exceed the power of the black 
hole since the disk has much smaller internal resistance. But in this case, even 
the black hole's power is only a small faction of the total power, it
can still be important in practice if the energy from the black hole is very
clean and not mixed with the energy from the disk \cite{bla99}.

In conclusions, for the simple model considered in the paper we have shown 
that the power of the disk dominates over the power of the black hole. This
suggests that the Blandford-Znajek mechanism may be less efficient for
extraction of energy from black hole with a thin disk. However, due to 
many simplifications we have made for the model, improvements are required
for getting more thorough understanding of the Blandford-Znajek mechanism.
Especially the generalization to the case of thick disk deserves considerations.

\acknowledgments{I acknowledge Bohdan Paczy\'nski, Jeremy Goodman, Richard Gott,
David Meier, Hyun Kyu Lee, Gerry Brown, Ralph Wijers, and Jiri Bicak for many helpful
discussions. I am very grateful to the referee Roger Blandford for many
valuable comments and suggestions. This work was supported by NASA Astrophysical
Theory Grant NAG5-2796.}


\begin{figure}
\caption{The variation of the ratio of disk's EMF, ${\cal E}_D$, to the
black hole's EMF, ${\cal E}_H$,
with respect to the parameter $a/M$. The curves are plotted from $a/M=0.1$
to $a/M=0.99$ for different values of $n$: $1, 2$, and $3$ [see the labels at
the left end of each curve; the surface current density in the disk is
$J\propto r^{-n}$]. The long dashed line represents ${\cal E}_D/{\cal E}_H=1$, 
the short dashed line represents ${\cal E}_D/{\cal E}_H=1/3$ which is the critical 
value for the most efficient case with $Z_A=Z_H$ (where $Z_H$ is
the electric resistance of the black hole, $Z_A$ is the electric resistance of
the remote astrophysical load). The parameter ${\cal E}_D/{\cal E}_H$
measures the relative importance of the power of the disk to
the power of the black hole [see Eq.~(\ref{pdh}) in the text and the 
discussions below it]. If ${\cal E}_D/{\cal E}_H>1$,
the power of the disk always dominates the power of the black hole. 
If ${\cal E}_D/{\cal E}_H>1/3$, the power of the disk dominates 
the power of the black hole in the most efficient case with $Z_A=Z_H$.}
\label{figure1}
\end{figure}

\begin{figure}
\caption{The ratio of the disk's power $P_D$ to the black hole's power $P_H$
for the case with $n=2$ and $Z_A=Z_H$. The vertical dashed line shows the position
(where $a/M=0.54$) for $P_H=I{\cal E}_H-I^2 Z_H=0$. For $a/M<0.54$, the black hole's
power $P_H$ is negative; for $a/M>0.54$, the black hole's power $P_H$ is positive
but less than the disk's power $P_D$. The shaded region ($0\le P_D/P_H<1$) shows 
the case when the Blandford-Znajek mechanism is efficient, i.e. when the black hole's
power dominates the disk's power. In the unshaded region, the disk's power dominates 
the black hole's power, the Blandford-Znajek mechanism is less efficient. 
None of the models is in the shaded zone, i.e. the B-Z mechanism is
less efficient in all models considered in the paper.}
\label{figure2}
\end{figure}

\begin{figure}
\caption{The variation of the ratio of the disk's torque, $T_D$, to the 
black hole's torque, $T_H$, with respect to the parameter $a/M$. The 
curves are plotted from $a/M=0.1$ to $a/M=0.99$ for different values of $n$: 
$1, 2$,and $3$ (see the labels 
at the end of each curve). The dashed line represents $T_D/T_H=1$. 
If $T_D/T_H>1$, the torque of the disk dominates the torque of the black hole;
if $T_D/T_H<1$, the torque of the black hole dominates the torque of the disk.}
\label{figure3}
\end{figure}

\begin{figure}
\caption{The left diagram shows the distribution of the magnetic field $B_H$ on
the black hole's horizon for $n=1$ and different values of $a/M$. 
The right diagram shows the distribution of the normal component of the
magnetic field $B_D$ on the disk for $n=1$ and different values of $a/M$,
drawn from $r=r_{ms}(a)$ to $r=8M$. The inner edge of the disk is at $r=r_{ms}(a)$,
as denoted with the thick dots at the left end of each curve.}
\label{figure4}
\end{figure}

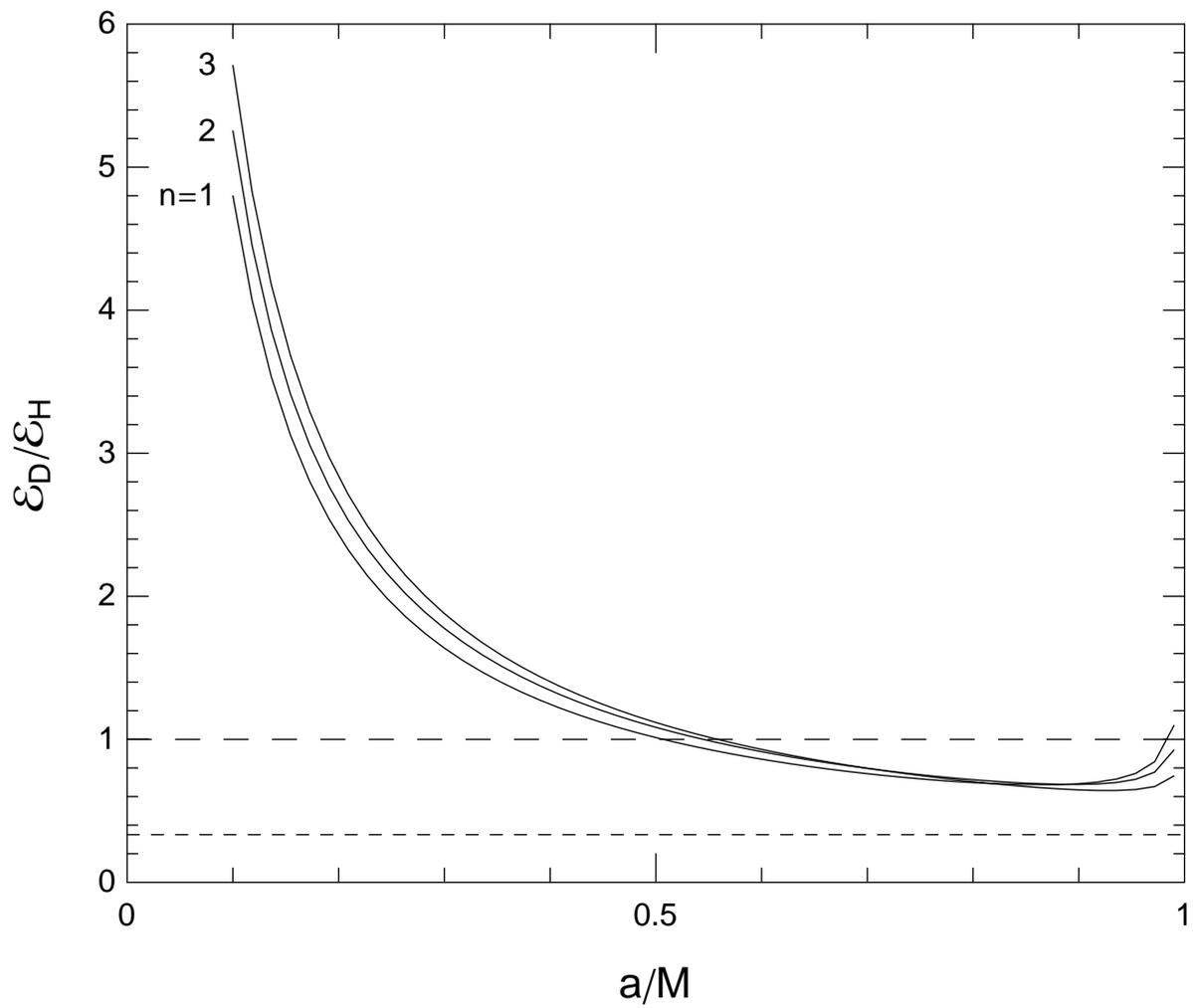
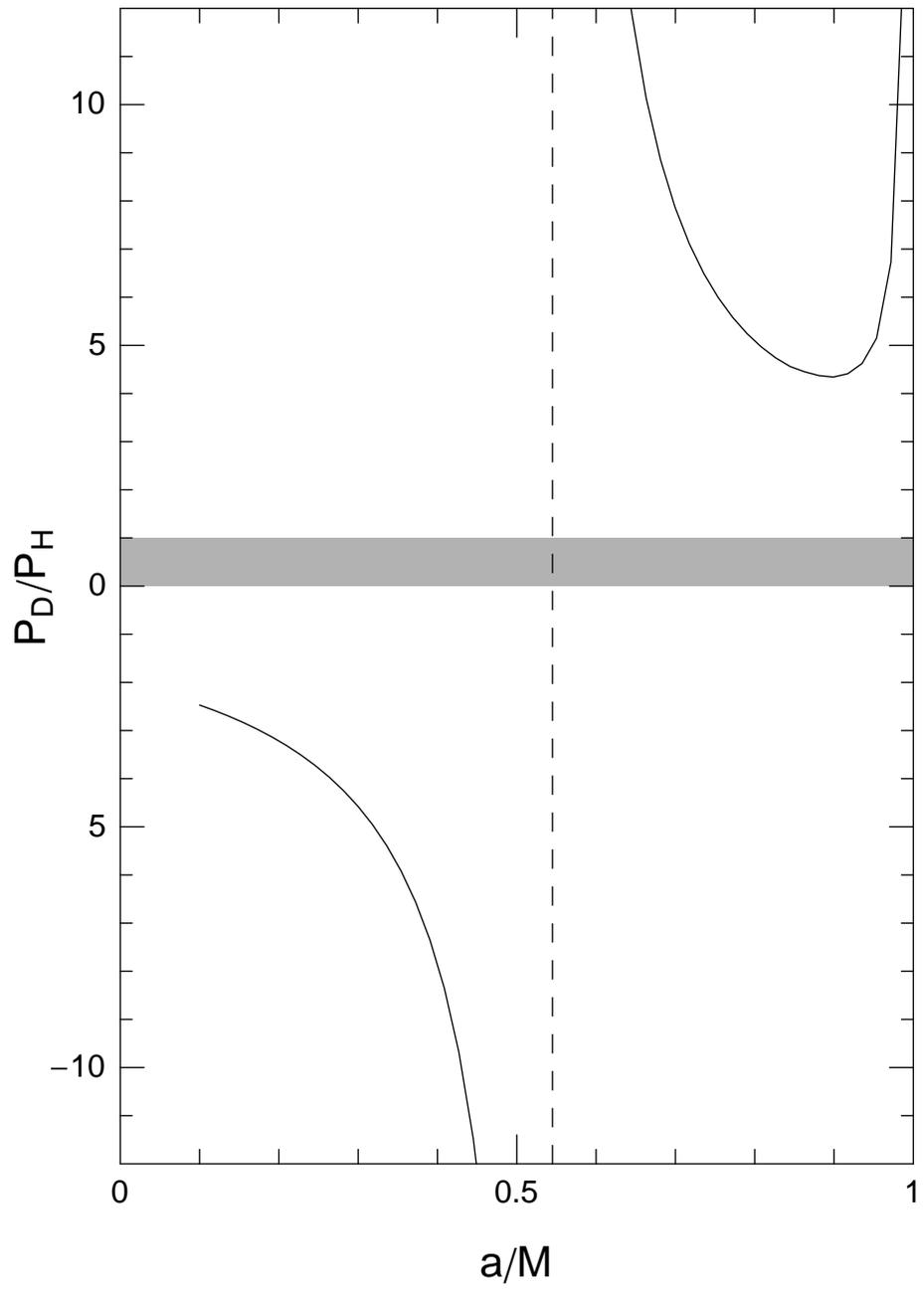
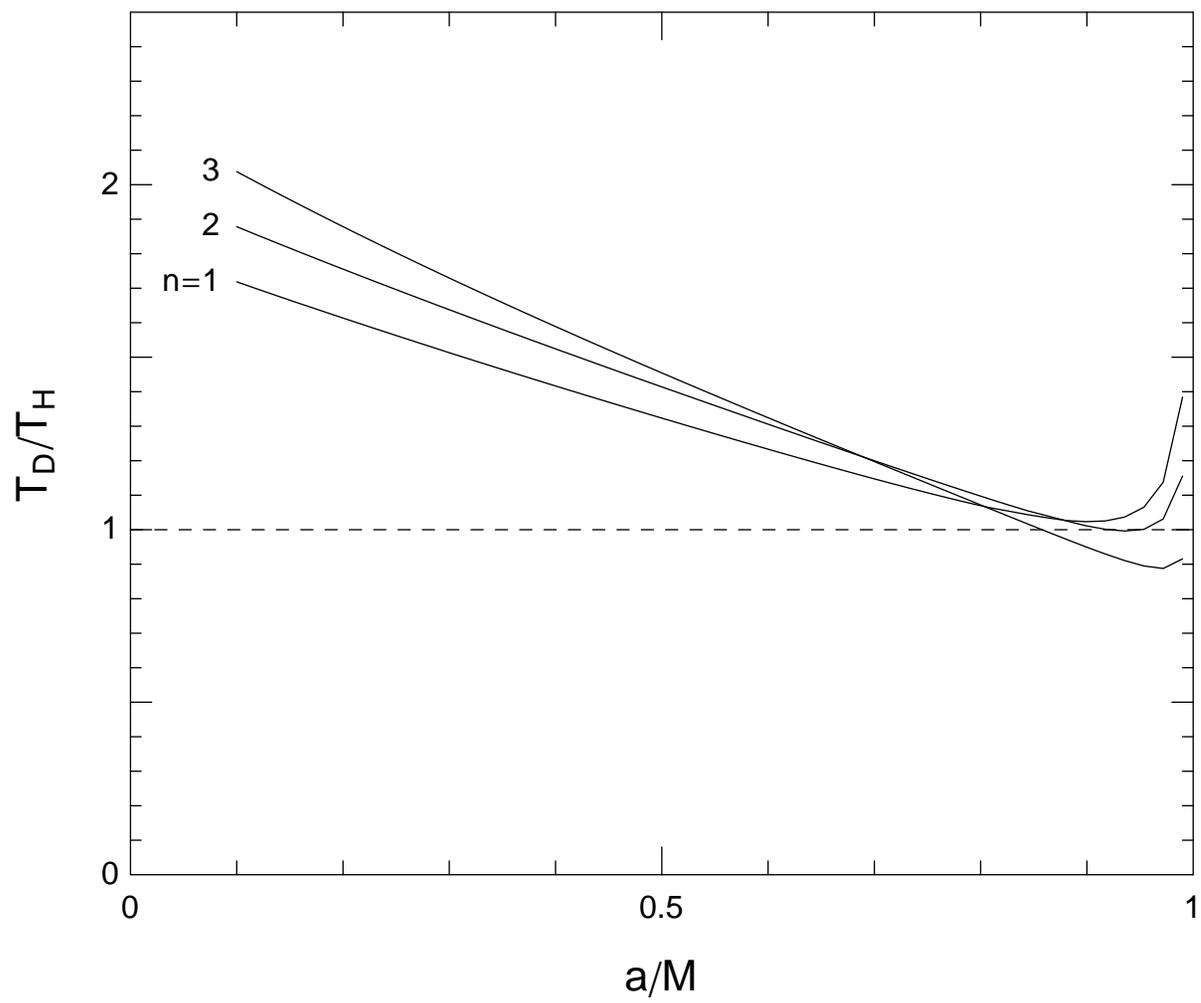
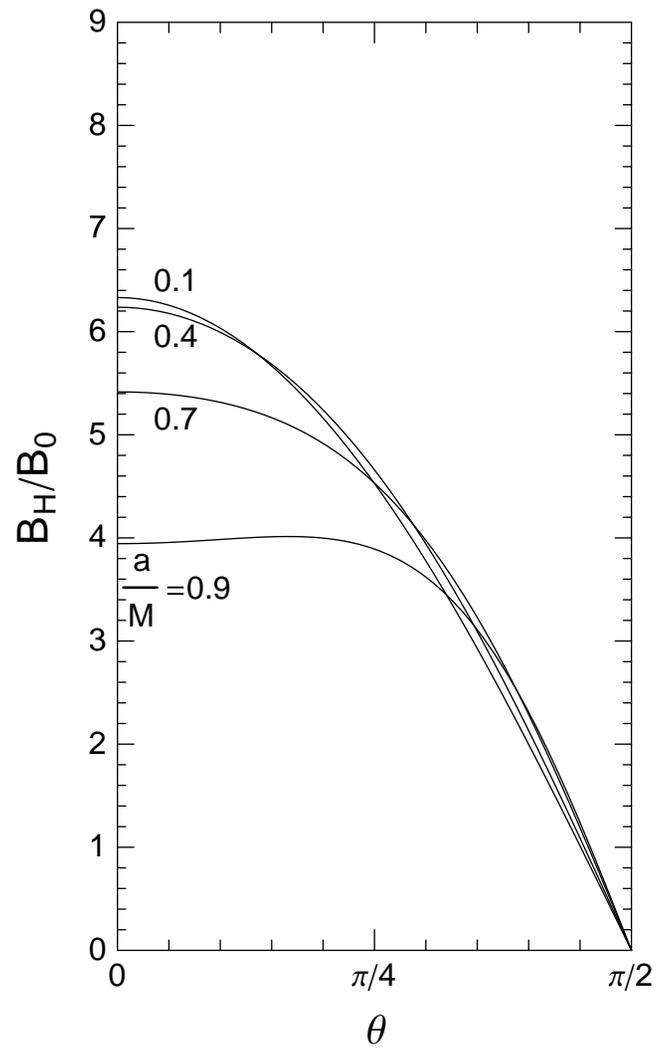
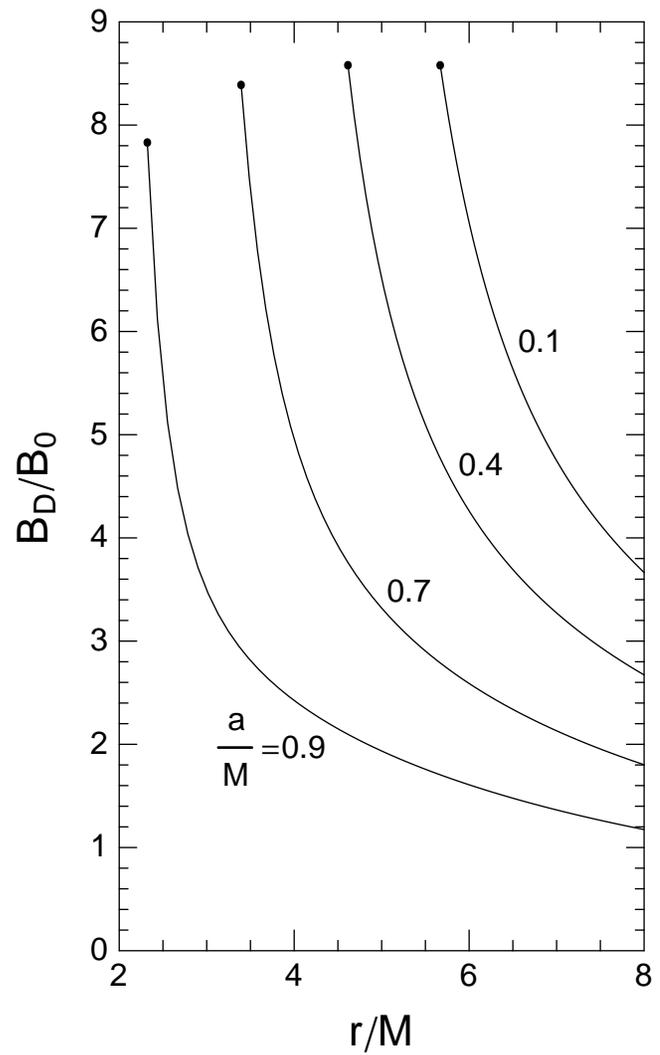
\end{document}